\documentclass[a4paper]{article}
\usepackage{epsfig,amssymb,amsmath,cite}

\def\beq{\begin{eqnarray}}
\def\eeq{\end{eqnarray}}
\def\bsp{\begin{split}}
\def\esp{\end{split}}

\def\diag{\mathrm{diag}}

\newcommand{\mb}[1]{{\mathbb #1}}
\newcommand{\mc}[1]{{\cal #1}}

\begin{document}

\title{\textbf{Brane-world Singularities}}
\author{\textbf{Alan A. Coley}$^1$\thanks{aac@mathstat.dal.ca}~ and \textbf{Sigbj\o rn Hervik}$^2$\thanks{%
S.Hervik@damtp.cam.ac.uk} \\
\\
${}^1$Department of Mathematics and Statistics,\\
Dalhousie University,\\
Halifax, Nova Scotia, Canada  B3H 3J5
\vspace{.5cm}\\
${}^2$DAMTP, Centre for Mathematical Sciences,\\
Cambridge University, \\
Wilberforce Rd.,\\
Cambridge CB3 0WA, UK}
\date{\today}
\maketitle

\begin{abstract}
We study the behavior of spatially homogeneous
brane-worlds close to the initial singularity in the presence of
 both local and nonlocal stresses. It is found that the
singularity in these brane-worlds can be locally either
isotropic or anisotropic. We then  investigate
the Weyl curvature conjecture, according to which some measure of the Weyl
curvature is related to a gravitational
entropy. In particular, we study the Weyl curvature conjecture
on the brane with respect to the dimensionless ratio of
 the Weyl invariant and the Ricci square and
 the measure proposed by Gr{\o}n and Hervik. We also argue that
 the Weyl curvature conjecture should be formulated  on brane
 (i.e., in the four-dimensional context).
\end{abstract}

\section{Introduction}
Recently there has been great interest in brane-world cosmological
models  \cite{Randall:1999vf,Randall:1999ee}, particularly in an
 attempt to understand the dynamics of the universe at early times. Brane-world models have a
 different qualitative behavior than their general-relativistic counterpart
 \cite{BDL,MSS:2001,Campos:2001pa,Toporensky:2001hi},
 especially at high energies when the energy density of the matter is larger than the brane tension and  the behavior
 deviates significantly from the classical case. However, in spite of this
 interest, the initial singularity seems to be very little understood
 in these brane-world models. The earliest investigations of the initial singularity,
 which used only isotropic fluids as a source of matter, suggest a matter-dominated isotropic singularity
 \cite{BDL,Coley1,Coley2,vdh}. However, later work using anisotropic stresses indicate that the initial singularity
 need not necessarily
 be isotropic \cite{BH}. These issues are also linked to whether or not the brane-worlds exhibit a
 chaotic singularity (see e.g. \cite{jenam}).

There are a number of important issues that need to be addressed in
early universe cosmology. Although inflation is a possible causal
mechanism for homogenization and isotropization, there is a
problem in that the initial conditions must be
sufficiently smooth in order for inflation to subsequently take
place \cite{Coley1}. This problem of initial conditions
 in inflation might be alleviated
by a regime of chaotic mixing \cite{YST,LMP,CSS}; consequently the
possibility of  chaotic behaviour in brane-world models  is of
interest
 \cite{Coley2,jenam}. However, more importantly,
 an isotropic singularity   in
brane world cosmology might provide for  the
 necessary sufficiently smooth initial
conditions, and might in turn be explained by entropy arguments
and the second law of thermodynamics \cite{penrose}. The study of
the initial singularity in cosmology may consequently be closely
related to the notion of ``gravitational entropy'' and the arrow
of time \cite{davies74,davies83}. A more precise statement of  this
is the Weyl Curvature Conjecture (WCC) according to which the Weyl
tensor should be related to the entropy of the
gravitational field \cite{penrose,wcc1,wcc2,wcc3}.

Therefore, in this paper we will investigate the WCC for
brane-world models. In order to do this we have to carefully study
the singularity of the brane-world models in the
presence of general fluids with anisotropic stresses (such an
investigation is lacking at present). An
important conclusion from this investigation is, as we will show,
that there exist two local past attractors for these brane-worlds; one
isotropic past attractor and one anisotropic past attractor. For
the isotropic singularity, the measure $P$ defined by the ratio of
the Weyl invariant and the Ricci square, \beq
P^2=\frac{C^{\alpha\beta\gamma\delta}C_{\alpha\beta\gamma\delta}}{R^{\mu\nu}R_{\mu\nu}},
\eeq
 decreases to zero as the singularity is approached, consistent
 with the WCC. The measure $\mc{S}$  \cite{wcc1,wcc2}, defined on the brane by
\beq \mc{S}=\sqrt{h}P, \eeq
 behaves according to the WCC for both of these singularities.

 We note that the WCC for brane-worlds
 involves the five-dimensional Weyl tensor; additional terms arising from the projection of the five-dimensional
 Weyl tensor onto the brane should be included in the brane-world scenario. However, we will argue that
  the WCC should be formulated in the four-dimensional context.

\section{The equations of motion}
We  consider spatially homogeneous brane (or branes where the
spatial derivatives can be neglected). The evolution equations on
the brane are as follows \cite{Maartens:2001jx,Maartens:2000fg}.
The Friedmann equation,
\begin{eqnarray}
H^2&=&\frac \Lambda 3+\frac 16\sigma ^{\mu \nu }\sigma _{\mu \nu }-\frac 16%
{}^{(3)}\mathcal{R}+\frac{\kappa ^2}3 \rho \nonumber \\
&+&\frac{\kappa ^2}{6\lambda}\left[ \rho^2  -\frac 3{2}\pi _{\mu
\nu }\pi ^{\mu \nu }\right]
+\frac{2%
\mathcal{U}}{\kappa ^2\lambda },  \label{constraint}
\end{eqnarray}
the shear propagation equations,
\begin{eqnarray}
\dot{\sigma}_{\langle\mu \nu \rangle}+\Theta \sigma _{\mu \nu
}&=&\kappa ^2\pi
_{\mu \nu }-{}^{(3)}\mathcal{R}_{\langle\mu \nu \rangle}\nonumber \\  &+& %
\frac{\kappa ^2}{2\lambda }\left[ -(\rho +3p)\pi _{\mu \nu }+\pi
_{\alpha
\langle\mu }\pi _{~\nu \rangle}^\alpha \right] +\frac 6{\kappa ^2\lambda }\mathcal{P}%
_{\mu \nu }, \label{CMeq}
\end{eqnarray}
Raychaudhuri's equation ($\Theta=3H$),
\begin{eqnarray}
&&\dot{\Theta}+\frac 13\Theta ^2+\sigma ^{\mu \nu }\sigma _{\mu
\nu }+\frac
12\kappa ^2(\rho +3p)-\Lambda =  \nonumber \\
&&-\frac 1{2\lambda \kappa ^2}\left[ \kappa ^4(2\rho ^2+3\rho p)+12\mathcal{U%
}\right],
\end{eqnarray}
the dark energy propagation equation,
\begin{eqnarray}
&&\dot{\mathcal{U}}+\frac 43\Theta \mathcal{U}+\sigma ^{\mu \nu }\mathcal{P}%
_{\mu \nu }=  \nonumber \\
&&\frac{\kappa ^4}{12}\left[ 3\pi ^{\mu \nu }\dot{\pi}_{\mu \nu
}+3(\rho +p)\sigma ^{\mu \nu }\pi _{\mu \nu }+\Theta \pi ^{\mu \nu
}\pi _{\mu \nu }-\sigma ^{\mu \nu }\pi _{\alpha \mu }\pi _\nu
^{~\alpha }\right].
\end{eqnarray}
Here, $H$ is the Hubble parameter; $\sigma_{\mu\nu}$ is the shear
tensor; ${}^{(3)}\mathcal{R}_{\mu\nu}$ is the 3-curvature; $\rho$
the energy-density; $p$ the isotropic pressure; $\pi_{\mu\nu}$ the
anisotropic stress tensor; $\mathcal{U}$ and
$\mathcal{P}_{\mu\nu}$  are the nonlocal dark energy and the bulk
graviton stress tensor, respectively. The nonlocal energy-flux
$\mc{Q}_{\mu}$ has to vanish due to the shear constraint equation.
Note that there are no propagation equations for the nonlocal bulk
graviton stress tensor, $\mathcal{P}_{\mu\nu}$.

For a FRW brane near the initial singularity, the Friedmann
equation gives \beq H^2\propto \rho^2, \eeq and thus $H, ~\rho
\propto t^{-1}$. The three-volume, $v$, goes as $v\propto
t^{1/\gamma}$. We will primarily be interested in spatially
homogeneous and spatially flat branes. The FRW universe corresponds to an
equilibrium point of the
equations of motion when written as a dynamical system, and hence
may serve as  a
local past attractor.
The behavior of the brane-worlds close to these
equilibrium points,
which  have the important property that
 $H\propto t^{-1}$,
 are reduced to linear algebra by linearizing
the set of equations with respect to the equilibrium point.

In the following we will assume that the brane-world contain two
types of fluids; an isotropic fluid obeying the barotropic
equation of state $p_i=(\gamma-1)\rho_i$, and a radiation type of
fluid causing the  anisotropic stresses $\pi_{\mu\nu}$. The fluids
separately obey the energy-momentum conservation equations \beq
\dot\rho_i+\gamma\Theta\rho_i &=& 0, \nonumber \\
\dot\rho_r+\frac 43\Theta\rho_r +\sigma^{\mu\nu}\pi_{\mu\nu}&=& 0.
\label{eq:rhos} \eeq

\section{Isotropic or anisotropic initial singularity?}
In an earlier paper \cite{BH} it was shown that the initial
singularity for a magnetic brane-world  could be either locally
isotropic or anisotropic. The question we address here
is: What happens in the case with more general types of stresses?
Will the same behavior persist? Will there exist one anisotropic
and one isotropic past attractor for more general brane-worlds?

In contrast to the general-relativistic models, the brane-worlds
will in general be matter-dominated as we approach the initial
singularity. The question whether the singularity is anisotropic
or isotropic reduces to the question of which fluid dominates at the
initial singularity. If the anisotropic fluid dominates initially,
the fluid will typically drive the initial singularity to an
anisotropic state. If, on the other hand, the isotropic fluid
dominates, the universe will be driven to an isotropic FRW state.

\subsection{Isotropic singularity}
Let us first assume that the
initial singularity is isotropic; hence we have a FRW singularity
with $H,~\rho_i \propto t^{-1}$ so that the three-volume  expands
as a power-law $v\propto t^{1/\gamma}$. Moreover, isotropy implies
that the anisotropic stresses and fluid have to be small in the
sense that $\pi_{\mu\nu},~ \sigma_{\mu\nu},~\rho_r \ll \rho_i$.

To check the stability of the isotropic singularity we first
investigate the stability of the ratio of $\rho_r/\rho_i$. For the
isotropic singularity to be stable for this combination of fluids, we must have
$\rho_r/\rho_i\rightarrow 0$ as $t\rightarrow 0$. Using eq.
(\ref{eq:rhos}), we get the following requirement for the
isotropic singularity to be past stable (using also that
$\pi_{\mu\nu},~ \sigma_{\mu\nu},~\rho_r \ll \rho_i$) \beq \gamma>
4/3. \eeq It is quite easy to understand this requirement
intuitively. Close to isotropy the anisotropic fluid behaves as
isotropic radiation. If the isotropic fluid has $\gamma < 4/3$,
then the anisotropic fluid will be dominant initially. This would,
however, drive the universe into an anisotropic state and hence
the isotropic FRW would be unstable. If, on the other hand, the
isotropic fluid is stiffer than radiation, then the isotropic
fluid  dominates initially and the isotropic
singularity is stable. Close to the isotropic singularity  the shear will
 behave as a $\gamma=1$ fluid and will therefore also be
suppressed near the initial singularity. We note that $\gamma> 4/3$
for scalar field matter close to the singularity (which acts effectively
as stiff matter with $\gamma_{eff}=2$).

The only thing we have assumed in the above heuristic analysis is
that the dominant term in the Friedmann equation is the
$\rho_i^2$-term. The dark-energy $\mc{U}$ behaves close to
isotropy as classical radiation and will therefore be sub-dominant
to the anisotropic radiation fluid (which come with quadratic
terms in the Friedmann equation). The nonlocal stresses
$\mc{P}_{\mu\nu}$ have the same source as the dark-energy (they
both are different components of the Weyl tensor in the bulk) and
hence one should expect that the nonlocal stresses are
sub-dominant as well. The above stability analysis is therefore
quite general and encompasses many different types of brane-world
scenarios.

To find the asymptotic behavior of the shear we have to specialize
to a certain type of anisotropic fluid. We assume that the
anisotropic stresses are of the form
$\pi_{\mu\nu}=C_{\mu\nu}\rho_r$, where $C_{\mu\nu}$ is a
matrix of constants (or slowly varying functions). This choice includes the magnetic field as a special case
and may give us a useful indication of the asymptotic behavior of the
shear in the presence of stresses.

Consider the isotropic singularity with $\gamma >4/3$. Assuming
that $\mc{P}_{\mu\nu}$ behaves approximately as the dark-energy
term we can neglect the nonlocal stresses on the right-hand-side
of the shear propagation equations. To lowest order  the equations
are \beq \dot{\sigma}_{\langle\mu \nu \rangle}+\Theta \sigma _{\mu
\nu } \propto \rho_i\rho_r\propto t^{-1-\frac 4{3\gamma}}. \eeq
This equation can be integrated to give to lowest order \beq
\sigma_{\mu\nu}=\mc{O}\left(t^{-\frac{4}{3\gamma}}\right). \eeq

\subsection{Anisotropic singularity}
This case is more difficult than the isotropic one since the
anisotropic singularity is almost entirely determined by the
anisotropic matter. Also, the nonlocal stresses come into play and
are difficult to deal with in full generality. The main reason for
this is that from a brane point of view, there are no evolution
equations for $\mc{P}_{\mu\nu}$.  In principle, $\mc{P}_{\mu\nu}$ should arise as
a solution to the full five-dimensional theory, but we shall
 assume $\mc{P}_{\mu\nu}=0$ in the
  specific example (equilibrium point) discussed below.
However, by a
continuity argument we can expand the results to
$\mc{P}_{\mu\nu}=\mc{U}D_{\mu\nu}$, where $D_{\mu\nu}D^{\mu\nu}$
is sufficiently small.

We start out with the assumption that there exists an anisotropic
equilibrium point in the past. We can then give a criterion
that must be fulfilled if the equilibrium point is past stable.
For an important class in which there exists
such an equilibrium point, we shall show that
the equilibrium point is past stable if and only if this
criterion is satisfied.

We assume that there exists an anisotropic equilibrium point such
that $\pi_{\mu\nu},~\sigma_{\mu\nu}\propto t^{-1}$, and $\rho_i\ll
\rho_r$. We also assume the asymptotic values of $\rho_r$ and $\Theta$ are
$\rho_r=r/t$ and $\Theta=\theta_0/t$, respectively, where $r$ and $\theta_0$ are
constants. This implies that we have the asymptotic value \beq
\frac{\sigma^{\mu\nu}\pi_{\mu\nu}}{\rho_r^2}\approx
\mathrm{constant}\equiv D \eeq Using eq. (\ref{eq:rhos}) we get
the following asymptotic relation 
\beq -1+\frac 43\theta_0 +D r=0.
\label{eq:thetadrrelation}\eeq 
Consider the ratio $R=\rho_i/
\rho_r$. For the equilibrium point to be past stable  $\dot{R}>0$,
so using eqs. (\ref{eq:rhos}) and (\ref{eq:thetadrrelation}), we
obtain 
\beq \theta_0\gamma <1. \label{eq:thetagamma}\eeq 
The constant
$\theta_0$  is related to the volume expansion rate as $v\propto
t^{\theta_0}$, so eq. (\ref{eq:thetagamma}) gives an upper limit on
the exponent on the power-law behavior of the volume close to the
equilibrium point. Note that if $\theta_0<1/2$, then this inequality
is automatically satisfied for all $\gamma\in [0,2]$.

So far we have only analysed the stability in the $R$
direction. In order for the equilibrium point to be past stable,
it has to be stable in the remaining variables as well.
Let us now provide an example which is past stable in the
other variables. The example is a two-parameter set of
different matter configurations which interpolates between the
magnetic brane-worlds and a  brane-world with two isotropic
fluids. Only
 eq.(\ref{eq:thetagamma}) then remains to be satisfied.

Given an anisotropic stress in the diagonal form
\beq
\pi_{\mu\nu}=\rho_r\diag\left(0,x+\sqrt{3}y,x-\sqrt{3}y,-2x\right),
\eeq where $x,y$ are constants and we assume that the nonlocal
stresses vanish, $\mc{P}_{\mu\nu}=0$. We will also assume that the
dominant energy condition \beq
|p_j|=\left|\frac{\rho_r}{3}+\pi_{jj}\right|\leq \rho_r \eeq is
satisfied for the fluid $\rho_r$. This implies that $(x,y)$ are
restricted to a compact subset of $\mb{R}^2$.

We define the following constants \beq
\Delta &=& 2(x^2+y^2)+x^3-3xy^2 \nonumber \\
\Sigma^2 &=& (2x+x^2-y^2)^2+(2y-2xy)^2 \nonumber \\
L^2&=&2\left[2+9(x^2+y^2)\right]\Delta \nonumber \\
a &=& 36\Delta^2+8\Sigma^2+6L^2\nonumber \\
B &=& \sqrt{(a+L^2)^2-8a(\Sigma^2+L^2)}-L^2. \eeq Then there exist
a past equilibrium point given by the asymptotic values \beq
\Theta &= &\frac{3(a-B)}{4a}\frac 1t,\nonumber \\
\sigma_+ &=& \frac{(2x+x^2-y^2)B}{6\Delta a}\frac 1t,\nonumber \\
\sigma_- &=& \frac{(2y-2xy)B}{6\Delta a}\frac 1t,\nonumber \\
\frac{\kappa^2}{\lambda}\rho_r^2 &=& \frac{B(B+a)}{12\Delta a^2}\frac 1{t^2},\nonumber \\
\rho_i&=& 0,\nonumber \\
\frac{2\mc{U}}{\kappa^2\lambda} &=&
\frac{9\Delta^2(a-B)^2-4B^2\Sigma^2-2\Delta
B(B+a)\left[1-9(x^2+y^2)\right]}{144\Delta^2 a^2}\frac{1}{t^2}.
\eeq Here are $\sigma_{\pm}$ related to $\sigma_{\mu\nu}$ via \beq
\sigma_{\mu\nu}=\diag\left(0,\sigma_++\sqrt{3}\sigma_-,\sigma_+-\sqrt{3}\sigma_-,-2\sigma_+\right).
\eeq It can be shown that these asymptotic solutions are past
attractors within the subset $\mc{P}_{\mu\nu}=0$, \emph{provided
that} inequality (\ref{eq:thetagamma}) holds. Further, it can be
shown that $\theta_0<3/4$ for   nonzero $x,y$.

Using a continuity argument\footnote{The argument uses the fact
that finding the equilibrium points of a dynamical system is
basically a question of finding solutions to an algebraic set of
equations. Also, the stability analysis for these equilibrium
points depends on the eigenvalues of a set of linear equations
which are continuous functions of the $C_{\mu\nu}$ and
$D_{\mu\nu}$. } we can show that \emph{there also exist a past
attractor for models with nonlocal anisotropic stresses of  type
$\mc{P}_{\mu\nu}=\mc{U} D_{\mu\nu}$ where $D_{\mu\nu}$ is sufficiently small.}
Hence, there is a class of models with
$\mc{P}_{\mu\nu}\neq 0$ which have an anisotropic past attractor.
How large this class is, and if this anisotropic past attractor
exists for generic brane-worlds, needs further  work.\footnote{The
paper \cite{SST} considers brane dynamics with nonlocal stresses.
However, the brane-world does not contain an additional
anisotropic fluid and thus there are no anisotropic  past
attractors.}

\section{The Weyl Curvature Conjecture}
We have seen how different brane-world scenarios have  two
different past attractors. Here we will explore both of these
possibilities further and check whether the brane-worlds behave
according to the WCC (see \cite{wcc1,wcc2} for a review).

For the isotropic singularity we need to know how  the Weyl
tensor approaches the asymptotic values. In the previous section
we found that $\sigma_{\mu\nu}\propto t^{-4/3\gamma}$ (recall that
$\gamma>4/3$ in this case). The dominant term in the Weyl
curvature invariant will therefore be $\left(\dot\sigma_{\mu\nu}\right)^2$.
Hence, \beq
\left(C^{\alpha\beta\gamma\delta}C_{\alpha\beta\gamma\delta}\right)_I
= t^{-4}\cdot \mc{O}\left(t^{\frac{2}{3\gamma}(3\gamma-4)}\right).
\eeq The Ricci square behaves as $t^{-4}$, thus \beq
P_I &=&\mc{O}\left(t^{\frac{1}{3\gamma}(3\gamma-4)}\right),\nonumber \\
\mc{S}_I &=&\mc{O}\left(t^{\frac{1}{3\gamma}(3\gamma-1)}\right).
\eeq Hence, the isotropic singularity behaves according to both
the $P$-version and the $\mc{S}$-version of the WCC in this case.

For the anisotropic singularity, the behavior is easier to
establish. The Weyl curvature behaves according to \beq
\left(C^{\alpha\beta\gamma\delta}C_{\alpha\beta\gamma\delta}\right)_A
= t^{-4}. \eeq The Ricci tensor evolves similarly, so
\footnote{Although $P_A$ is asymptotically  a constant, its derivative
can be  either sign asymptotically.}
\beq
P_A &=& \mathrm{constant}, \nonumber \\
\mc{S}_A&\propto & t^{\theta_0}. \eeq 
Recall that $\theta_0$ obeys the
bound $\theta_0\gamma <1$, and the explicit value of $\theta_0$
depends very much on the matter content. In all of the cases we
have explored here, $\theta_0 <3/4$.

We can conclude that the measure $\mc{S}$ behaves according
to the WCC for both the anisotropic and the isotropic singularity.

\section{A four-dimensional or five-dimensional WCC?}
So far we have considered only the four-dimensional version of the
WCC, seen from the brane point of view. However, a question that
arises  for these brane-worlds is whether one should consider the
WCC from a five-dimensional perspective or a four-dimensional
perspective. Even a
four-dimensional measure of
``gravitational entropy'' should have a part arising from the
projection of the five-dimensional Weyl tensor onto the brane.
Hence the gravitational entropy on the brane will have one part
corresponding to  the four-dimensional Weyl tensor, and a part
from the remaining components of the projection of the
five-dimensional Weyl tensor. These components are\footnote{It
should be noted that all these entities should be defined in the
bulk in the limit close to the brane. The Weyl tensor itself
experiences a delta function singularity at the brane
\cite{mannheim}.} \beq
\mc{E}_{\mu\nu}={}^{(5)}C_{ABCD}n^An^Cg^B_{~\mu}g^D_{~\nu}, \\
\mc{B}_{\mu\nu\alpha}={}^{(5)}C_{ABCD}n^Dg^C_{~\alpha}g^A_{~\mu}g^B_{~\nu},
\eeq where $n^A$ is a unit vector orthogonal to the brane, and
$g^A_{~\mu}$ is the projection tensor onto the brane (see
\cite{SSM1,SSM2}). These nonlocal terms give rise to  ``dark
energies'' and stresses that do not come from standard model
particles. More explicitly, the tensor $\mc{E}_{\mu\nu}$ can be
decomposed as \beq
\mc{E}_{\mu\nu}=-\frac{6}{\lambda\kappa^2}\left[\mc{U}\left(u_{\mu}u_{\nu}+\frac
13h_{\mu\nu}\right)+\mc{P}_{\mu\nu}+2\mc{Q}_{(\mu}u_{\nu)}\right],
\eeq and thus is, up to a constant factor, formally equivalent to the energy-momentum
tensor of a radiation fluid with energy-density $\mc{U}$,
anisotropic stress $\mc{P}_{\mu\nu}$, and energy-flux
$\mc{Q}_{\mu}$.
 The entropy of particles is calculable from standard thermodynamics, and hence there should also correspond an entropy
 due to the dark energies. These terms are purely gravitational effects but behaves very much like ordinary matter when it comes to the evolution of four-dimensional spacetime. Actually, the notion of gravitational entropy might even be more easily understandable for these terms than for the more mysterious Weyl tensor.

The five-dimensional Weyl curvature invariant can be written
as \beq
{}^{(5)}C^{ABCD}C_{ABCD}={}^{(4)}C^{\alpha\beta\gamma\delta}C_{\alpha\beta\gamma\delta}+6\mc{E}^{\mu\nu}\mc{E}_{\mu\nu}+4\mc{B}^{\mu\nu\alpha}\mc{B}_{\mu\nu\alpha}.
\eeq The tensor $\mc{B}_{\mu\nu\alpha}$ can be expressed in terms
of the exterior curvature of the hypersurfaces as \cite{SSM2} \beq
\mc{B}_{\mu\nu\alpha}=2\nabla_{[\mu}K_{\nu]\alpha}+\frac
23\left(\nabla_{\beta}K^{\beta}_{~[\mu}-\nabla_{[\mu}K\right)g_{\nu]\alpha}
\eeq where the covariant derivative, $\nabla_{\mu}$, is the
covariant derivative associated with the metric on the brane,
$g_{\alpha\beta}$. Further, from the junction conditions and the
$\mb{Z}_2$ symmetry, we can relate the exterior curvature close to
the brane and the energy-momentum tensor on the brane, \beq
K^+_{\mu\nu}=-K^-_{\mu\nu}=-\frac
16\kappa^2_5\tilde{\lambda}g_{\mu\nu}-\frac
12\kappa_5^2\left(T_{\mu\nu}-\frac 13g_{\mu\nu}T\right). \eeq In
our case, where the brane is spatially homogeneous, the components
$\mc{B}_{\mu\nu\alpha}$ can be replaced with time-derivatives of
the matter on the brane.

There are different versions of the WCC that  can be investigated.

\subsection{The WCC: Modified four-dimensional version}
First, we will consider the four-dimensional version where we take
into account the extra terms from the projection of the
five-dimensional Weyl tensor. Only the four-dimensional versions
of the tensors shall be used.

For the isotropic singularity, we will consider a FRW brane for
which exact five-dimensional solutions are known. In this case,
\beq
\left(\mc{B}^{\mu\nu\alpha}\mc{B}_{\mu\nu\alpha}\right)_{FRW}=0,
\eeq using the junction conditions and energy-momentum
conservation of the fluid on the brane. The term
$\mc{E}^{\mu\nu}\mc{E}_{\mu\nu}$ is, in general, given by \beq
\mc{E}^{\mu\nu}\mc{E}_{\mu\nu}=
\frac{36}{\lambda^2\kappa^4}\left(\frac
43\mc{U}^2+\mc{P}^{\mu\nu}\mc{P}_{\mu\nu}-2\mc{Q}^\mu\mc{Q}_{\mu}\right).
\eeq Hence, for the FRW brane\footnote{For an anisotropic brane
with shear degrees of freedom, in addition to this mode  we get the
usual shear modes from the four-dimensional Weyl invariant,
${}^{(4)}C^{\alpha\beta\gamma\delta}C_{\alpha\beta\gamma\delta}$,
found earlier.}, \beq
{}^{(4)}C^{\alpha\beta\gamma\delta}C_{\alpha\beta\gamma\delta}+6\mc{E}^{\mu\nu}\mc{E}_{\mu\nu}+4\mc{B}^{\mu\nu\alpha}\mc{B}_{\mu\nu\alpha}\propto
\mc{U}^2. \eeq For $\gamma\geq 2/3$ the Ricci square on the brane
diverges as $\rho^2$, so \beq
P_{FRW} &\propto & t^{\frac{2}{3\gamma}(3\gamma-2)}\\
\mc{S}_{FRW} &\propto &t^{\frac{1}{3\gamma}(6\gamma-1)}. \eeq For
the anisotropic singularity, we get a similar behavior as in the
pure brane case: \beq
\left({}^{(5)}C^{ABCD}C_{ABCD}\right)_A &\propto & t^{-4},\nonumber \\
P_A &=& \mathrm{constant}, \nonumber \\
\mc{S}_A&\propto & t^{\theta_0}. \eeq

\subsection{The WCC: The five-dimensional version}
The five-dimensional curvature tensors experience
$\delta$-function singularities at the location of the brane and
thus care is needed in this case. Let
us therefore consider the case where the volume under
consideration is entirely in the bulk but very close to the brane.

The measure
${}^{(5)}P={}^{(5)}C^{ABCD}C_{ABCD}/{}^{(5)}R^{MN}R_{MN}$ can be
used to investigate the five-dimensional WCC in the bulk.  The
bulk, which is  taken to be an Einstein space with negative
curvature\footnote{Recently  a
scalar field in the bulk has also been included, but these spacetimes have only
been investigated at a perturbative level.}, has
${}^{(5)}R_{AB}\propto \Lambda_B {}^{(5)}g_{AB}$. On the brane,
the five-dimensional Ricci tensor also experiences a
$\delta$-function singularity. Hence, it is of crucial importance
where one wants to consider the WCC. If we consider ${}^{(5)}P$ in
the bulk, close to the brane, we can use the known exact solution
for the FRW brane.

The FRW-brane can be embedded in an ambient Schwarzschild-AdS
space \cite{BDEL} \beq
ds^2=-f(R)dT^2+\frac{dR^2}{f(R)}+R^2\gamma_{ij}dx^idx^j,\quad
f(R)=1-\frac{U}{R^2}-\frac{\Lambda_B}{6}R^2, \eeq where
$\gamma_{ij}$ is the 3-dimensional spatial metric encompassing
   the fifth dimension. For simplicity, we have also assumed the
   closed FRW model. The induced metric on the brane is
\beq
ds^2=-d\tau^2+R(\tau)^2\left[\frac{dr^2}{1-r^2}+r^2(d\theta^2+\sin^2\theta
d\phi^2)\right], \eeq where $\tau$ is the proper time of the
brane, and $R(\tau)$ describes
   the motion of the brane; i.e. $R(\tau)$ can be identified as the
   scale factor $a(t)$. We can then easily find the Weyl curvature
   invariant:
\beq \left({}^{(5)}C^{ABCD}C_{ABCD}\right)_{FRW,bulk}\propto
R^{-8}. \eeq Hence, \beq
\left({}^{(5)}P\right)_{FRW,bulk}&\propto & R^{-4},\nonumber \\
\left({}^{(5)}\mc{S}\right)_{FRW,bulk}&\propto & R^{-1}. \eeq
Thus, both measures diverges as we approach the initial
   singularity.

For the anisotropic singularity, unfortunately
there are not many exact solutions known to study the possible behavior. However, there are exact
Kasner solutions \cite{Frolov}, and we can use these
solutions to obtain \beq
\left({}^{(5)}\mc{S}\right)_{A,bulk} \rightarrow \infty, \text{ as
} t\rightarrow 0. \eeq
Therefore we  conclude that the measures investigated
here diverge badly in the bulk as we approach the initial
singularity. Thus the WCC as formulated above is not valid in the
bulk.

This is perhaps not very surprising since
the physics in the bulk is very different than the physics on the
brane; indeed, the matter in the
bulk need not even obey the weak
energy condition. Moreover,
matter is confined to the  four-dimensional brane, and the physical energy conditions are only satisfied on the brane.
Therefore, we conclude that the {\it WCC should be formulated in the four-dimensional
context}.


\section{Conclusion}
 In this paper we have presented a
  thorough investigation of the initial
  singularity in brane-world cosmological models. 
  It was shown that for a  class of spatially homogeneous brane-worlds with
  anisotropic stresses, both local and nonlocal, the brane-worlds
  could have either an isotropic singularity or an anisotropic
  singularity.

We also discussed the Weyl Curvature Conjecture,
according to which the gravitational entropy is related to the
Weyl curvature, in the context of the newly proposed
brane-worlds. We found that
the isotropic equilibrium point is typical in the brane-world scenario.

For both the isotropic and the anisotropic  singularity the Weyl curvature
invariant,
$C^{\alpha\beta\gamma\delta}C_{\alpha\beta\gamma\delta}$, the Weyl
to Ricci ratio,
$P^2=C^{\alpha\beta\gamma\delta}C_{\alpha\beta\gamma\delta}/R^{\mu\nu}R_{\mu\nu}$,
and the measure $\mc{S}=\sqrt{h}P$ were investigated. The measure
$\mc{S}$ behaved according to the WCC in both cases,
while $P$ showed generic increasing behavior only for the
isotropic  singularity.  Hence
$P$ and $\mc{S}$ can be distinguished as measures of gravitational
  entropy.

However, for the brane-worlds a ``gravitational entropy'' may also
have additional terms arising from the higher-dimensional bulk.
The nonlocal terms, $\mc{U}$ and $\mc{P}_{\mu\nu}$, are
projections of the five-dimensional Weyl tensor and  behaves very
much like a fluid upon the cosmological evolution of the universe.
Hence, these terms are, in some sense, a direct manifestation of
the Weyl tensor in terms of dark energies and stresses. These
terms should therefore be incorporated into a gravitational
entropy, one way or another. However, it seems as if the WCC can
only be formulated on the brane where it can be seen upon as the
``shadow'' of the five-dimensional theory.

Finally, we remark that  brane-worlds may serve as an
interesting arena for studying the WCC. Actually, the brane-worlds
may serve as a scenario were gravitational entropy and its
relation to the Weyl tensor may be more easily understood.
\section*{Acknowledgments}
SH acknowledges the kind hospitality of Dalhousie University where
most of this work was done. A travel grant from Churchill College,
Cambridge, is also gratefully acknowledged. AAC was funded by the
Natural Sciences and Engineering Research Council of Canada and SH
was funded by the Research Council of Norway and an Isaac Newton
Studentship.

\end{document}